\documentclass[onecolumn,preprintnumbers,11pt,amsmath,amssymb,nofootinbib]{revtex4}
\usepackage{graphicx}
\def\be{\begin{equation}}
\def\ee{\end{equation}}
\def\bea{\begin{eqnarray}}
\def\eea{\end{eqnarray}}
\def\lf{\left (}
\def\rt{\right )}

\begin{document}




\title{
Brane-World Motion in Compact Dimensions}

\author{Brian Greene$^{1,2}$, Janna Levin$^{1,3}$, and Maulik Parikh$^{4}$}

\affiliation{$^{1}$Institute for Strings, Cosmology and Astroparticle Physics,
  Columbia University, New York, NY 10027, USA \\
$^{2}$Departments of Physics and Mathematics, Columbia University,
  New York, NY 10027, USA\\
  $^{3}$Department of Physics and Astronomy, Barnard College of
  Columbia University, New York, NY 10027, USA \\
$^{4}$Department of Physics and Beyond: Center for Fundamental Concepts in Science \\
Arizona State University, Tempe, AZ 85287, USA} 

\begin{abstract}
\begin{center}
{\bf Abstract}
\end{center}
\noindent
The topology of extra dimensions can break global Lorentz invariance,
singling out a globally preferred frame even in flat spacetime. 
Through experiments that probe
global topology, an observer can determine her
state of motion with respect to the preferred frame. This scenario is
realized if we live on a brane universe moving through a flat space with
compact extra dimensions.
We identify three experimental effects due to the motion of our
universe that one could potentially detect using gravitational
probes. One of
these relates to the peculiar properties of the twin paradox in multiply-connected spacetimes.
Another relies on the fact that the Kaluza-Klein modes of any bulk field are
sensitive to boundary conditions. A third concerns the modification to
the Newtonian potential on a moving brane. Remarkably, we find that
even small extra dimensions are detectable by brane observers if the
brane is moving sufficiently fast.
\end{abstract}

\maketitle

\section{Introduction}
\label{introduction}

In higher-dimensional theories of gravity, brane worlds define a
surface on which our Standard Model fields display conventional
(3+1)-dimensional physics. 
In many brane-world models, such as the one proposed by
ADD~\cite{ADD}, the extra dimensions are compact with a flat metric.
Even though Lorentz transformations remain local isometries,
the compactness of the extra dimensions
violates global Lorentz invariance. In particular, when the extra
dimensions are compactified by taking a quotient of $\mathbb{R}^n$ (as
in toroidal compactification), the direction of identification picks
out a globally preferred frame, despite the Minkowski metric.
Consequently it is meaningful to speak of the
brane's absolute motion; relativism of motion is lost. Moreover, an observer on
a brane universe moving at constant velocity can perform globally
sensitive experiments to determine the universe's velocity through the
extra dimension.

In this paper, we present three potentially measurable
effects of brane motion: (a) gravitons created at the brane take two different periods of time to travel around the compact dimension; (b) the Kaluza-Klein
tower splits into a tower of left-moving and a tower of right-moving
states, with different spacings; and (c) the Newtonian potential is
modified in such a way as to effectively magnify the size of the extra
dimension, $L$, by the Lorentz factor, $\gamma$, with the $4d$ effective gravitational
coupling, $G_N$, now related to the $5d$ coupling, $G$, through 
\be
G_N=\frac{G}{\gamma L} \; .
\ee
Although possibly difficult to
observe, if these effects were detectable, they could deliver measures
of the size of the
extra dimension and the velocity of our brane universe through the extra dimension.

In \S \ref{preferredframe} we review the peculiar features of special
relativity on cylindrical Minkowski space, in particular the existence
of a globally preferred frame. In \S \ref{movingbranes} we realize the
topological breaking of Lorentz symmetry in terms of brane worlds
embedded in flat compact extra dimensions. We point out several
different kinds of effects of brane motion.
Time-delayed interactions and graviton return times are described in \S \ref{fireworks}.
We obtain the split in the Kaluza-Klein spectrum 
in \S \ref{KK}. Low-energy deviations from $1/r^2$ gravity that depend
on the boost as well as the size of the circle are presented
in \S \ref{newtonianpotential}. In \S \ref{discussion} we discuss our underlying assumptions and indicate some directions for further work.

\section{Special Relativity on a Cylinder}
\label{preferredframe}

The principle of relativity asserts the equivalence of all inertial observers. 
In special relativity, as in Galilean relativity, there are no preferred observers, and
only relative inertial motion has meaning. For special relativity, this fundamental tenet is consistent with the statement that spacetime is an $\mathbb{R}^n$ manifold with the Minkowski metric; Lorentz transformations that take one inertial observer into another leave the Minkowski line element invariant. 
However, the equivalence of all inertial observers breaks down when spacetime has non-trivial topology. This happens even in flat space. Consider the simple example of a two-dimensional cylindrical spacetime, the product of a circle with the time axis. A cylinder of course is intrinsically flat: the metric in every coordinate patch is precisely the Minkowski metric, the Riemann tensor vanishes, and parallel lines do not meet. Yet, despite being locally identical to ordinary Minkowski space, cylindrical Minkowski space has some unusual, even surprising, properties. 

For example, because the worldlines of two inertial observers can intersect multiple times as
the observers circumnagivate the multiply-connected dimension, the twin paradox
takes on a new and more subtle character \cite{bransstewart,peters,barrowlevin,bl2,luminettwin}. 
On a cylinder neither twin needs to have experienced a period of acceleration in order to reunite with the other twin. Both twins move on geodesics. From the absence of noninertial forces, each twin knows he is inertial and could be tempted to conclude that 
the laws of special relativity should therefore apply. Each would then think
of himself as at rest and the other twin as moving, and therefore as
younger. Of course, when they 
meet they cannot both be right.

The resolution of this unaccelerated version of the twin paradox is that, notwithstanding the Minkowski metric, Lorentz symmetry is broken globally by the topology. This is easy to see on the covering space. 
The cylinder can be obtained from the $\mathbb{R}^n$ universal covering space
through a quotient along a spacelike
direction. There is then a preferred frame: observers whose worldlines are orthogonal to the axis of identification are special. The existence of a preferred set of observers seems to violate Lorentz
invariance and that is because Lorentz invariance \emph{is} violated,
globally though not locally, by the topological identification. And generally it is the case
that when we quotient a spacetime by a discrete isometry,
we break the isometry group globally.\footnote{Unless the quotient
  is by an element of the center of the isometry group, as happens in
  the ``elliptic" identification of de Sitter space \cite{dSZ2}.} In the case of the cylinder, spacetime is still locally
Minkowski space and therefore appears to have Lorentz
symmetry. But the identification has glued one coordinate to itself at
a fixed time, thereby singling out as special those observers whose spatial coordinate
coincides with that along the circle, and rendering all other time slices inequivalent. (Put another way, there is a unique spacelike Killing vector whose integral curves form closed orbits.) What distinguishes the
twins then are their 
speeds with respect to the preferred
observer. The twin with the higher speed comes back younger (when the twins have the same speed but opposite velocities, they return with the same age, despite their relative motion). In general, the age difference between the twins can be
resolved quantitatively by evaluating the proper times of their worldlines in the
covering space \cite{barrowlevin}. 

More precisely, consider two-dimensional Minkowski space with topology $\mathbb{R}^2$ and local line element
\be
ds^2 = - dt^2 + dy^2 \; .
\ee
We wish to compactify along the $y$ direction so that the
topology becomes $S^1 \times \mathbb{R}$. To cover the circle we can
choose either a single-valued but discontinuous coordinate or a
multi-valued but continuous coordinate. Choosing the latter, we
identify the coordinates via
\be
 \left ( \begin{array}{ccc} t \\ y  \end{array} \right ) \sim 
  \left ( \begin{array}{ccc} t \\ y + L \end{array} \right ) \; , \label{preferid}
\ee
%
where $L$ is the circumference of the circle. By
virtue of having selected the $y$ direction, this identification picks out 
preferred observers, those whose
worldlines are orthogonal to the $y$-axis. For these observers, space is 
a circle of circumference $L$.

The existence of a preferred frame makes it meaningful to speak of absolute 
speed. Indeed, by performing experiments that probe the global
topology of the space, an inertial observer can unambiguously
determine whether she is moving.
Here is a simple experiment an inertial observer could perform to
determine absolute motion: Send a probe around the cylinder. 
Suppose an inertial observer
pair-produces two particles
that move in opposite directions along the extra spatial direction.
Momentum conservation requires
that the two particles have opposite velocities as measured by the observer who produced them.
When the preferred observer, $O$, does this experiment, he finds that
particles moving at speed $s$ intercept his worldline again at the
same time, $L/s$.

\begin{figure}[hbtp]
\centering
\includegraphics[angle=0,width=120mm]{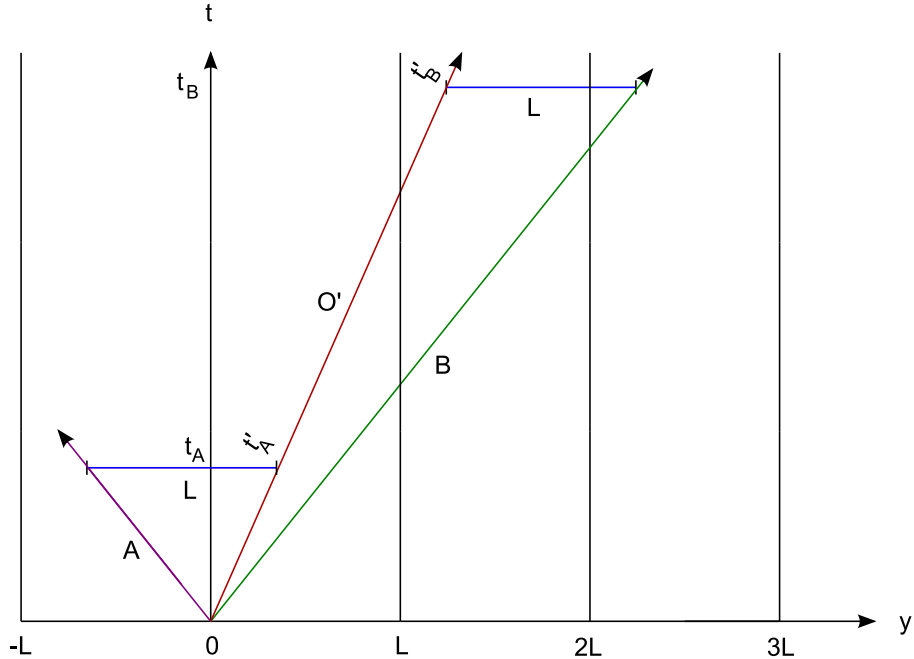}
\caption{Spacetime diagram of the universal covering space. The size
  of the space in the preferred coordinates $(t, y)$ is $L$. The
  worldlines of a non-preferred observer, $O^\prime$, and two particles, $A$
  and $B$, are indicated. Particle $A$ leaves the observer at the
  origin and returns with period $t_A$ in preferred coordinates and
  $t'_A$ in the observer's proper time. Particle $B$ returns with a
  longer period.}
 \centering
\label{fig:1}
\end{figure}

But consider the same experiment performed by a boosted observer, $O^\prime$. This observer records a quite different result: the particles return to him at {\em separate} times since the left-moving probe and the right-moving
probe intercept $O^\prime$s world line at different events (Fig.\ \ref{fig:1}). Suppose the probes are sent out in opposite directions at speed $s'$, as measured by $O^\prime$, who is moving to the right with 
speed $\beta$. Let $A$ be the particle moving to the left of
the observer, and let $B$ be the particle moving to the right. Then
the amount of preferred time that has elapsed when the probes meet
$O^\prime$'s worldline again are
\be
t_A = {L \over s'} \left ({1 - s' \beta \over 1 - \beta^2}\right )
\qquad t_B = {L \over s'}\left ({1 + s' \beta \over 1 - \beta^2}\right ) \; .
\ee
If the particles are massless $(s'=1)$, the
return times, expressed in terms of $O^\prime$'s proper time, are
\be
t_{A}^\prime={L \over \gamma (1 + \beta)} \qquad
t_{B}^\prime= {L \over \gamma(1 - \beta)} \; ,
\ee
where the Lorentz factor, $\gamma$, is $(1-\beta^2)^{-1/2}$.

The preferred observer has $\beta = 0$, so the return periods
are the same. 
In general, any inertial observer can deduce his 
speed with respect to the
preferred frame from the return times. Let $\tau_{\rm short} =
t_{A}^\prime$ be the shorter time and $\tau_{\rm long} = t_{B}^\prime$ the longer time, as measured by $O^\prime$. Then
\be
\beta = {1 \over s'} {\tau_{\rm long} - \tau_{\rm short} \over \tau_{\rm long} +
  \tau_{\rm short}} \; , \label{vobs}
\ee
where $s'$ is the speed of the probes in the frame of the observer who sent them. In other words, $O^\prime$ can perform the experiment of sending, say, two
photons ($s' = 1$) in opposite directions --- and upon receipt of those same
photons --- conclude that he was in uniform motion, without reference to
any other observer. In one and
only one frame --- the preferred frame --- do the two returning
photons reach the observer simultaneously.\footnote{This resembles the Sagnac
effect in which photons traveling down the different arms of a
rotating ring-shaped interferometer are subject to different path
lengths and transit times, a phenomenon that is important for global positioning system
satellites in rotational orbit about the Earth.} 
The observer can determine his speed (though not his velocity) with respect to the preferred frame from the difference in photon return times.

A corollary of this is that a family of parallel, moving, inertial observers, at rest with respect to each other, cannot globally synchronize their clocks using Einstein clock synchronization. 
This is because Einstein clock synchronization methods (sending light signals
back and forth) fail to give a unique synchronization for the non-preferred observers: on the cylinder, there is more than one way to send a
light signal back, and synchronizing in one direction gives a different result
from synchronizing in the other direction. Einstein clock synchronization requires that the travel times of the incoming and outgoing light signals be the same, but this is only true for a pair of preferred observers.

Furthermore, the natural coordinates of moving observers have
discontinuities in time as well as in space. Of course, since a circle
is not homeomorphic to a segment of the real line, even the preferred
observer cannot avoid coordinate discontinuities. But these are just
the usual discontinuities in which the spatial coordinate goes from
$L$ back to $0$. In the moving coordinate system, however, the
discontinuities also afflict the time coordinate.
Consider an observer $O^\prime$ moving
with velocity $\beta$ relative to the preferred frame.
Let $t'$ be his proper time, and let $y'$ coordinatize the spacelike direction orthogonal
to $\partial_{t'}$ in the $y-t$ plane. The coordinates $(t,y)$ and $(t',y')$
are related by a Lorentz transformation 
as follows:
\be
 \left ( \begin{array}{ccc} t' \\  y'  \end{array} \right ) = \Lambda
 \left ( \begin{array}{ccc} t \\ y  \end{array} \right )
\quad , \quad 
\Lambda = \left ( \begin{array}{ccc} \gamma & -\gamma \beta \\ -
   \gamma \beta & \gamma  \end{array} 
\right ) \; . \label{lorentztrans}
\ee
Acting on both sides of (\ref{preferid}) with the Lorentz matrix,
$\Lambda$, gives the identification in the primed coordinates:
\be
 \left ( \begin{array}{ccc} t' \\ y' \end{array} \right ) \sim 
  \left ( \begin{array}{ccc} t' - \gamma \beta L \\ y' + \gamma L
    \end{array} \right ) \; . \label{nonpreferid}
\ee
If a single-valued time coordinate is used, then, at some
arbitrary point in space,
the time coordinate is forced to jump by a finite amount;
evidently, Cartesian coordinates set up by $O^\prime$ suffer discontinuities in both space and time.\footnote{Indeed, this is a familiar phenomenon: 
a similar thing happens
across the International Date Line. The world-volume of the surface of
the Earth has topology $S^2 \times \mathbb{R}$, of course. But for the
purposes of assigning time zones the latitudes play no role; only
longitudes matter (neglecting the tilt of the Earth's axis of
rotation). So, as far as time zones go, the relevant topology
is actually $S^1 \times \mathbb{R}$, a cylinder. (More formally, time
zones do not extend to the poles; an $S^2$ minus the poles is
equivalent to an $S^1$, by a deformation retraction.) If we were to
use preferred time, geostationary clocks everywhere along the
Equatorial circle 
would show the same time. Although such
a coordinatization is possible, and even in some sense natural, on
Earth we prefer (not because of relativity, but purely out of convenience) to use a different time, one that tracks the motion of the Sun. Rather than assigning the same time
to the entire Equator, 
we choose a Sun-adapted coordinate system that is offset by an hour for every 15 degrees, relative to preferred time. Because such equal-time slices are tilted with respect to the preferred time-slices, they inevitably suffer temporal discontinuities,
making an International Date Line unavoidable.}

\section{Moving Brane Universes}
\label{movingbranes}

We have seen that the existence of nontrivial topology can break Lorentz invariance
globally by selecting a preferred frame. In particular, flat compact extra dimensions provide a concrete realization of this scenario. 
In the rest of this paper, we will consider brane worlds in flat compact extra dimensions. Since the non-trivial topology has broken global Lorentz invariance and picked out a preferred frame, it is meaningful to speak of the brane's velocity.
(Brane worlds are invoked here in order to localize the observer
within the extra dimensions, thereby ensuring that our
special-relativistic considerations --- which were framed in terms of
classical worldlines --- remain valid; had we used the conventional
Kaluza-Klein construction, our four-dimensional world would have been
smeared uniformly over the extra dimensions, rather than localized
within them.) We will find remarkably that --- just as an inertial
observer can use globally sensitive probes to determine his absolute
motion --- a brane observer can detect the motion of the brane universe through the extra dimension by using gravitational probes. Our discussion will be purely kinematical
and very generic; we will not refer to any specific brane model. The
only condition on the brane model is that the brane live in flat
compactified space \cite{ADD,DGP}; the Randall-Sundrum model
\cite{RS2} does not apply, not only because the extra dimension is not
compact but also because the curved background does not start with
Lorentz isometries.

Consider then a generic brane, to which Standard Model fields are
confined, living in a spacetime with flat compact extra
dimensions. The compact extra dimensions could be either large or
small, compared with the inverse of the cut-off scale; these lead to
different effects, 
three of which are discussed in the following subsections.

\subsection{Time-delayed fireworks}
\label{fireworks}

Let us assume at first 
that there is one large extra dimension with the topology
of a circle. Here  by ``large" we mean large compared to the inverse of the cut-off
scale. Bulk particles produced on the brane then have
characteristic wavelengths that are much smaller than the scale of the
extra dimensions; for instance, if the extra dimensions are of
millimeter size, and the UV cut-off on the brane is a TeV, then the
ratio of the de Broglie wavelength to the size of the extra dimension is
about $10^{-16}$. Such particles can effectively be described by
wavepackets moving on classical trajectories, i.e. on
worldlines. Hence when there are large extra dimensions, both the brane observer and the graviton probes can be treated as moving on classical worldlines, exactly as in the previous section.

This leads to the following effect. Imagine that, at the Large Hadron Collider, a collision of Standard
Model particles takes place in which massless bulk fields are excited. The
bulk particles travel around the extra dimensions on classical trajectories and return to the
brane. When they re-enter the accelerator (from the extra dimension),
they interact with the brane to produce Standard Model particles. If
all the energy is not deposited at once on the brane, the particles go
around additional times, depositing a little more energy on the brane with each collision,
in the form of ``fireworks" of Standard Model particles \cite{ADD,langloissorbo,langloisetal}. A particle experimentalist in Geneva would therefore observe a sequence of displaced
vertices. These interaction vertices would be equally separated in time with a period given by the
time taken by the bulk particles to circumnavigate the extra
dimension. 

However, if the brane happens to be moving through the extra dimension, then, by
the logic of the previous section, the vertices would appear with
\emph{two} sets of periods. By considering the difference between
these two periods, one could, via  (\ref{vobs}), deduce the speed of
our four-dimensional brane as it sails through the extra dimension.
More generally, the bulk gravitons could carry off momentum with
components tangential to the brane. 
For a brane with speed $\beta$,
the following equations are then obeyed in preferred coordinates:
\begin{eqnarray}
\beta t_{\rm long} + L & = & v_{y, {\rm long}} t_{\rm long} \nonumber \\
\beta t_{\rm short} - L & = & - v_{y, {\rm short}} t_{\rm short} \nonumber \\
x_{\rm long} & = &  v_{x, {\rm long}} t_{\rm long} \nonumber \\
x_{\rm short} & = &  v_{x, {\rm short}} t_{\rm short} \; .
\end{eqnarray}
Here $y$ is the compact direction while $x$ is a direction tangential to the brane. The subscripts ${\rm long}$ and ${\rm short}$ label whether the graviton took more or less time respectively to return to the brane. $x_{\rm long}$ and $x_{\rm short}$
are the spatial distances along the brane between the point of creation
of the pair of gravitons and the points of return of the gravitons.
These equations can be expressed in moving coordinates. The $x$ coordinates are
invariant since the boost is transverse to the $x$-direction, while
\begin{eqnarray}
t_{\rm long} & = & \gamma \tau_{\rm long} \nonumber \\
t_{\rm short} & = & \gamma \tau_{\rm short} \; .
\end{eqnarray}
The quantities $\tau_{\rm long}$, $\tau_{\rm
  short}$, $x_{\rm long}$, and $x_{\rm short}$ are all measurable by
an observer on the brane. We therefore have four equations in six unknowns. However, because the gravitons are massless, they also obey
\begin{eqnarray}
v_{x, {\rm long}}^2 + v_{y, {\rm long}}^2 & = & 1\nonumber \\
v_{x, {\rm short}}^2 + v_{y, {\rm short}}^2 & = & 1 \; .
\end{eqnarray}
Hence all six unknowns including the size, $L$, of the extra dimension, and the speed, $\beta$, of the brane can be deduced from the positions in spacetime of graviton interaction vertices. In particular, the brane speed is
\be
\beta = \frac{\left (\tau_{\rm long}^2 - \tau_{\rm short}^2
  \right ) - \left (x_{\rm long}^2 - x_{\rm short}^2 \right
  )}{\sqrt{\left ( \left (\tau_{\rm long} + \tau_{\rm short} \right
    )^2 - \left (x_{\rm long} - x_{\rm short} \right )^2 \right )
    \left ( \left (\tau_{\rm long} + \tau_{\rm short} \right )^2 -
    \left (x_{\rm long} +  x_{\rm short} \right )^2 \right )}} \; .
\ee
When there is no tangential motion ($x_{\rm long} = x_{\rm
  short} = 0$), this reduces to (\ref{vobs}). Remarkably then, using gravitational probes, the inhabitants of the brane can deduce the speed of their universe.
Of course if the graviton has a substantial $x$-component of velocity, the next
point of contact with the brane could well have moved outside of the
detector altogether, beyond Geneva, or even beyond the solar system. 
Perhaps another signal of brane motion could be a modification to the missing
energy. 

We have assumed that particles entering the extra dimension
return with probability one. In fact, the returning bulk particles have some interaction probability with the brane. If this probability is not close to unity, then bulk particles may
occasionally pass through the brane without producing any interaction vertices. This would then create a set of staggered displaced
vertices in which a certain fraction of vertices would be missing,
depending on the interaction probability. If there are
enough vertices though, one might still be able to determine the shortest period between
interactions. 

Unfortunately, this effect is likely to be very difficult to measure in practice because of the extreme smallness of the gravitational coupling. Moreover, another complication arises when there is more than one multiply-connected large extra
dimension. In that case, point particles leaving the brane will not
necessarily return to the brane. Consider a two-torus with modulus
$\tau = \tau_1 + i \tau_2$. A point-like particle will return to the brane only
if it is moving with slope
\be
{n \tau_2 \over n \tau_1 + m} \; ,
\ee
where $m$ and $n$ are integers. But particles moving with such velocities form a set of measure zero.
%
%
%
%

\subsection{Kaluza-Klein Modes}
\label{KK}


Now let us consider the opposite limit, in which the size of the extra dimension is comparable to the typical wavelength of the graviton. In this case, the graviton needs to be treated as a wave. Here too there are effects of brane motion, essentially because waves are sensitive to boundary conditions.
Consider a free massive scalar field $\phi(t, \vec{x},y)$. In
preferred coordinates, the field
obeys the Klein-Gordon equation:
\be
(\Box - m^2) \phi(t,\vec{x},y) = (- \partial_t^2 + \vec{\nabla}_x^2 +
\partial_y^2 - m^2) \phi(t, \vec{x},y)= 0 \; .
\ee
The mode functions are
\be
\phi_{k,q} \sim e^{-i \omega t} e^{i k x} e^{iqy} \; ,
\ee
where
\be
\omega^2 = k^2 + q^2 + m^2 \; .
\ee
Single-valuedness of the field under $y \sim y + L$ (Eq.\ (\ref{preferid})) implies that
\be
q = {2 \pi n \over L} \; , \label{preferdiscrete}
\ee
where $n$ is any integer. This is the usual familiar story.

Now consider the frame moving in the $+y$
direction with speed $\beta$. The local line element in the primed coordinates is the
usual flat space one and hence the wave operator is also the same. We
therefore write
\be
(\Box' - m^2) \phi(t',\vec{x}',y') = (- \partial_t'^2 +
\vec{\nabla}_x'^2 + \partial_y'^2 - m^2) \phi(t', \vec{x}',y')= 0 \; .
\ee
For the mode functions this means
\be
\phi_{k',q'} \sim e^{-i \omega' t'} e^{i k' x'} e^{iq'y'} \; ,
\ee
where again
\be
\omega'^2 = k'^2 + q'^2 + m^2 \; . \label{omegaprime2}
\ee
However, in primed coordinates, the identification is given by (\ref{nonpreferid}), which mixes space and time components. Single-valuedness of the field under this identification requires that
\be
e^{-i \omega' t'} e^{i k' x'} e^{iq'y'} = e^{-i \omega' (t'- \gamma
  \beta L)} e^{i k' x'} e^{iq'(y'+ \gamma L)} \; .
\ee
Hence
\be
\gamma \beta \omega' + \gamma q' = {2 \pi n \over L} = q \; .  \label{qLT}
\ee
This is just the inverse Lorentz transformation acting on the momentum. (There is also the corresponding equation for $\omega$, namely $\gamma \beta q' + \gamma \omega' = \omega$.)
We see that, in the non-preferred frame, boundary conditions discretize a linear combination of 
momentum and energy, in contrast to
(\ref{preferdiscrete}). Substituting for $q'$ into
(\ref{omegaprime2}), we find that
\be
\omega' =  \gamma \sqrt{k'^2 + m^2 + \left ( {2 \pi n \over L} \right
  )^{\! 2}}  - \beta {2 \pi n \gamma \over L} \; . \label{omegaprime}
\ee
For the preferred observer, the left- and right-moving modes enter symmetrically in the dispersion relation:
\be
\omega^2 = k^2 + m^2 + \lf \frac{2 \pi n}{L} \rt ^{\! 2}
\ee
In particular, the standard Kaluza-Klein tower has a two-fold degeneracy, with positive and negative $n$ having the same energy; this reflects the fact that left-moving and right-moving bulk modes are treated symmetrically when there is no brane motion. For massless fields with $|k| \ll 2 \pi |n|/ L$, the energy spectrum has a two-fold degeneracy with fixed spacing $\omega \approx 2 \pi |n|/L$. However, when there is brane motion, the degeneracy is lifted. From (\ref{omegaprime}) we have for low $|k|$ that
\begin{eqnarray}
\omega'\approx q'\approx & 
\frac{2 \pi n}{\gamma L(1+\beta)}\qquad & n > 0 \nonumber \\
\omega' \approx  q' \approx & 
\frac{2 \pi |n|}{\gamma L (1-\beta)}\qquad & n < 0
\end{eqnarray}
The Kaluza-Klein tower of states splits up into two interlaced
towers. The tower of left-moving states and the tower of right-moving
states have different spacings.
As $\beta \rightarrow 1$, the frequency of the right-moving states $\rightarrow 0$, and
therefore they become easier to excite, while the frequency of the left-moving states
$\rightarrow \infty$, and therefore they become harder to excite.

In terms of the corresponding wavelength for massless $k=0$ modes, 
\begin{eqnarray}
\lambda_n
=\frac{\gamma L}{n}(1+\beta)\qquad & n > 0 \nonumber \\
\lambda_n=\frac{\gamma L }{|n|}(1-\beta)\qquad & n < 0
\label{fig:lambda}
\end{eqnarray}
We have expressed the factors in this way to suggest that an observer on a moving brane perceives a {\it larger} extra dimension with length $\gamma L$. Indeed, (\ref{nonpreferid}) already indicates that, in order for the invariant interval between identified points to be unchanged, the identification along the spatial axis is larger in primed coordinates than in preferred coordinates.
This magnification of extra dimensions by a Lorentz factor
can be visualized as follows.
For the moving observer, the spatial axis is tilted with
respect to that of the preferred observer. Along this tilted axis, the
proper length of the extra dimension once it spirals around and
intersects the observer's worldline is indeed $\gamma L$, as
illustrated in Fig 2.
\begin{figure}[hbtp]
\centering
\includegraphics[angle=0,width=80mm]{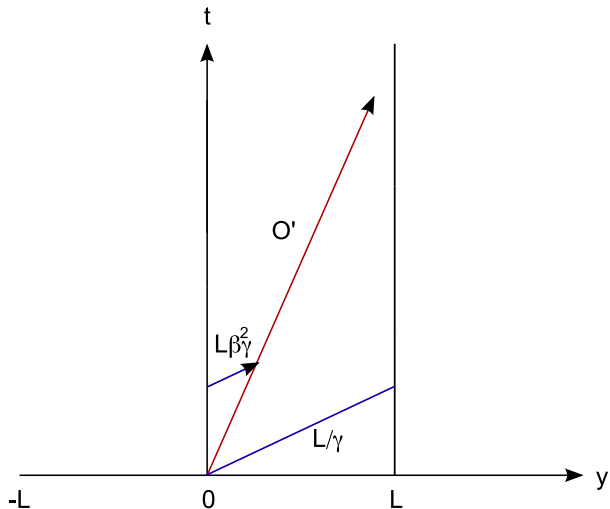}
\caption{The effective size of the extra dimension for the moving
  observer is $L/\gamma + L \beta^2 \gamma = \gamma L$.}
 \centering
\label{fig:2}
\end{figure}
In the next section, we will see that the combination $\gamma L$ is also the physically relevant scale in the modification of the Newtonian potential.

The factor of $(1\pm \beta)$ in (\ref{fig:lambda}) is similarly understood.
The right-moving standing wave that
corresponds to
the lowest eigenmode ($n=1$) can reconnect with its origin on the
brane only after it catches up to the brane and so extends over a distance
even larger than the brane measures. The left-moving standing wave
that corresponds to the lowest eigenmode ($n=-1$) can reconnect with its
origin on the brane when the brane closes the gap to meet it and so
extends over a distance
smaller than the size of the internal dimension that the brane
measures. 


\subsection{Newtonian potential on a moving brane}
\label{newtonianpotential}

Apart from signatures at accelerators, brane motion also affects the
Newtonian potential. Here we calculate the departure of the
four-dimensional gravitational potential from the Newtonian $1/r$
form, as seen by an observer on a moving brane. Perhaps surprisingly, 
the effect is potentially measurable even if the extra dimensions are
very small, as they are in standard Kaluza-Klein compactification,
provided that the brane is moving sufficiently quickly through the
extra dimensions.

We are interested in the Newtonian potential between two sources on the 
moving brane that are at rest with respect to each other.
We calculate the corrected gravitational potential in preferred five
dimensional coordinates, $X =(t,\vec{x},y)$, 
and transform the result. 
The locations of mass $m_1$ and mass $m_2$ are
\be
X_1=
\begin{pmatrix}
t_1\\
\vec x_1 \\
y_1=\beta t_1
\end{pmatrix}
\, , \, 
X_2=
\begin{pmatrix}
t_2\\
\vec x_2 \\
y_2=\beta t_2
\end{pmatrix} \; .
\ee
The graviton exchange occurs between masses separated in time 
by $\Delta t=t_1-t_2$ and on the
brane by $\vec r=\vec x_1-\vec x_2$. As both masses are located on the brane, 
they are
displaced from each other in the extra direction by $\Delta
y=y_1-y_2=\beta \Delta t$ according to preferred observers. Hence
\be
\Delta X=X_1-X_2=
\begin{pmatrix} 
\Delta t \\ \vec r \\ \beta \Delta t
\end{pmatrix} \; .
\ee
From the path integral
representation, $Z=e^{iW}$, the interaction potential is
computed from
\be
W=-\frac{1}{2} \int d^5X\int d^5\bar X \ T^{\mu \nu}(X)D_{\mu \nu,\lambda
  \sigma}(X-\bar X)T^{\lambda \sigma} (\bar X) \; ,
\label{Eq:W}
\ee
where $D_{\mu \nu, \lambda\sigma}$ is the graviton propagator. Since
all indices are fully contracted over we can express the resultant
scalar in terms of rest mass quantities and hereafter drop the tensor indices
on the propagator to treat the graviton as a massless scalar. The
contraction is simplest in the frame of the brane where the only non-zero
contribution to the energy momentum tensor is
\be
T^{t't'}=m_1\delta^3(\vec x'-\vec x_1')\delta(y')+m_2\delta^3(\vec
x'-\vec x_2')\delta(y') \; .
\ee
Writing
\be
W=-\frac{8\pi G}{2} \int d^5X\int d^5\bar X J(X)D_F(X-\bar X)J (\bar X)\; ,
\label{Eq:W2}
\ee
the factor of $8\pi G$ assures 
the correct Newtonian limit for canonically normalized fields and $G$
is the five-dimensional gravitational constant.
The corresponding scalar
source is
\begin{align}
J&=m_1\delta^3(\vec x'-\vec x_1')\delta(y')+m_2\delta^3(\vec
x'-\vec x_2')\delta(y') \nonumber \\
&=m_1\delta^3(\vec x-\vec x_1)\delta(y-\beta t_1)/\gamma +m_2\delta^3(\vec
x-\vec x_2)\delta(y-\beta t_2)/\gamma \; ,
\end{align}
where we have used $\delta(az)=\delta(z)/a$. The $m_1,m_2$ are rest masses
 and $D_F(X_1-X_2)$
is the Feynman propagator for the canonically normalized massless
scalar field that models the graviton exchange. We integrate over
the 8 delta functions and change the remaining integrations over $\int dX^0\int
d\bar X^0$ to $\int dT\int d(\Delta t)$. Then, not forgetting a factor of 2 from the $m_1m_2$ cross terms, the interaction piece in (\ref{Eq:W2}) becomes
\be
W_{\rm int}(r) =- T 8 \pi G \frac{m_1m_2}{\gamma^2} \int_{- \infty}^{+\infty} d (\Delta t) D_F(X_1-X_2) \; .
\ee

Without identification along $y$ (in our mostly plus metric signature), we would have
\be
D_F(\Delta X) =  - \int \frac{d^5 k}{(2 \pi)^5} \frac{e^{ik \cdot \Delta X}}{k^2 - i \epsilon} \; .
\ee
But because $y$ is compact, $k_y$ is discretized in units of $2 \pi/L$
so that $dk_y /(2 \pi) = 1/L$ and the fifth integral is replaced by a
sum. The Newtonian potential is therefore
\be
W_{\rm int}(r)  = T \frac{8 \pi G m_1 m_2}{(2 \pi)^4\gamma^2} \!
\sum_{n=-\infty}^{n=+\infty} \! \frac{1}{L} \! 
\int  \!  \frac{e^{ikr \cos \theta_k}  \, k^2 \sin \theta_k }
{-\omega^2 + k^2 +  \lf \frac{2 \pi n}{L} \rt^{\!2} - i \epsilon} dk \, d \theta_k \, d \phi_k\, d \omega \!
\int \! d(\Delta t )e^{-i(\omega  - 2 \pi n \beta/L) \Delta t} \; .
\ee
The $\Delta t$ integral gives $2\pi \delta(\omega - 2 \pi n
\beta/L)$. Since the $\omega$ integral runs from $0$ to $\infty$, the delta function eliminates the negative $n$ part of the sum, and integrates to $1/2$ when $n=0$.
After the angular integrations, we obtain
\be
W_{\rm int}(r)  = - i T \frac{8 \pi G m_1 m_2}{(2 \pi)^2 \gamma^2 L r} 
\lf \frac{1}{2} \int _{-\infty}^{+\infty} \frac{k \, e^{ikr}  }
{k^2 - i \epsilon} dk + \sum_{n=+1}^{n=+\infty}
\int_{-\infty}^{+\infty} \! \frac{k \, e^{ikr}  }
{k^2 +  \lf \frac{2 \pi n}{\gamma L} \rt^{\!2} - i \epsilon} dk \rt \; .
\ee
The integrals over $k$ can be written as closed contours in the upper half-plane giving
\be
W_{\rm int}(r)  =  
  T\frac{2 Gm_1m_2}{\gamma^2 L  r} \lf \frac{1}{2} + \sum_{n=+1}^{n=+\infty}
e^{-2\pi n r /( \gamma L)} \rt \; .
\ee
so that
\be
W_{\rm int}(r) = T \frac{Gm_1m_2}{ \gamma^2 L r} \frac{1 + e^{-2\pi r/(
    \gamma L)}} {1 - e^{-2\pi r/( \gamma L)}} \; . \label{newtonpotl}
\ee

Now $W$ is a scalar so $Z=e^{iW}$ is true in any frame. In brane
coordinates 
the particles are at rest and the energy is pure potential, $Z=e^{-iH'T'}=e^{-iV_{\rm
    brane}(r)T'}$. Since $T/T'=\gamma$ we have finally the
modified Newtonian potential as measured by observers at rest on the brane:
\be
V_{\rm brane}(r) = - \frac{Gm_1m_2}{ \gamma L r} \frac{1 + e^{-2\pi r/(
    \gamma L)}} {1 - e^{-2\pi r/(\gamma L)}} \; . \label{newtonpotl}
\ee
When $r \ll \gamma L$, $V_{\rm brane}(r)$ behaves like $-\frac{Gm_1 m_2}{\pi r^2}$, which is indeed the
five-dimensional Newtonian potential. On the other hand, at large
distances, $r \gg  \gamma L$, $V_{\rm brane}(r)$ goes as $-\frac{G m_1 m_2}{\gamma L
 r}$, which is just the four-dimensional potential, provided we define the effective
four-dimensional Newton's constant, $G_N$, to be
\be
G_N = \frac{G}{ \gamma L} \; . \label{newtonconstant}
\ee
Remarkably, in (\ref{newtonpotl}) and (\ref{newtonconstant}) the effective size of the compact space is $\gamma L$, rather than $L$. The extra dimensions appear {\em magnified}.
Corrections to the $1/r$ form of the gravitational potential arise
when $r$ becomes appreciable 
compared to $\gamma L$, rather than $L$.

This is a very interesting
result because it means that even if the extra
dimensions are very small, as in standard Kaluza-Klein
compactification on a torus, we could still detect them as deviations
from the four-dimensional Newtonian potential if, for some reason, our
brane were moving at an ultra-relativistic speed through the extra
dimensions. Through precision table-top gravity experiments, 
one might be able to extract the value of $\gamma L$ from data. In that case, the
other effects described earlier (difference in graviton return times, splitting of the Kaluza-Klein tower) would constitute nontrivial consistency checks between the measurements.

As a check, we can re-derive the result by doing the calculation
directly on the covering space. We will use the method of
images. Consider ordinary, infinite five-dimensional Minkowski space
covered
by two coordinate systems, $(t,\vec{x},y)$ and $(t',\vec{x}',y')$.
On the covering space, neither of these coordinates are in any way
pathological. Let the brane be aligned along $y'=0$ with an infinite
number of images displaced by $L$ along the {\em unprimed}
$y$-axis. This is the covering space picture of our
scenario.

On the covering space, the five-dimensional graviton propagator
$D_F(X)$ (dropping tensor indices again) is just
\be
D_F (X^\prime) = \frac{i}{8 \pi^2 \lf X^{\prime 2} \rt^{3/2}} \; ,
\ee
where
$X^{\prime 2} = -t^{\prime 2} + r^{\prime 2} + y^{\prime 2}$ is the
five-dimensional invariant distance-squared (see
e.g. \cite{barvinsky}). Then the Newtonian potential on the brane is
\be
V_{\rm brane}(r) = 4 \pi Gm_1m_2 \int_{- \infty}^{+\infty} d(\Delta t^\prime)
D(\Delta X^\prime) \; .
\ee
Here $D$ is related to $D_F$ by a sum over images
\be
D(\Delta X^\prime) = \sum_{n=-\infty}^{n=+\infty} D_F (\Delta X'_n) \; ,
\ee
where $\Delta X'_n$ is the difference between the source at the origin
and the images of the second mass. Since the identification in the primed coordinates is
\be
 \left ( \begin{array}{ccc} t' \\  \vec{x}' \\ y' \end{array} \right ) \sim 
  \left ( \begin{array}{ccc} t' - \gamma \beta n L \\  \vec{x}'  \\ y'
    + \gamma n L \end{array} \right ) \; , \label{nonpreferid2}
\ee
the images of $(t',\vec{r}', y'=0)$ are separated from the origin by
\be
\Delta X'_n = (\Delta t' - \gamma \beta n L, \vec{r}', \gamma n L) \; .
\ee
We can now readily evaluate the potential:
\begin{eqnarray}
V_{\rm brane}(r) 
& = & \frac{ Gm_1m_2}{2 \pi} \sum_{n=-\infty}^{n=+\infty}
\int_{-\infty}^{\infty} \frac{i d(\Delta t')}{\lf  -( \Delta t'- \gamma \beta n L)^2 + r^2 + (\gamma n L)^2 \rt^{3/2}} \nonumber \\
& = & -\frac{Gm_1m_2}{\pi}  \sum_{n=-\infty}^{n=+\infty}  \frac{1}{r^2 + (\gamma n L)^2}  \nonumber \\
& = & - \frac{Gm_1m_2} {\gamma L r} \frac{1 + e^{-2 \pi r/(\gamma L)}}{1-
  e^{-2 \pi r/(\gamma L)}} \; . \label{imagecharge}
\end{eqnarray}
To obtain the second equality we have Wick-rotated ($\tau = it$) the line of integration, and for the last line we used the identity
\be
\sum_{n = - \infty}^{n=+ \infty} \frac{1}{x^2 + \lf \frac{n \pi}{a} \rt^{\! 2}} = \frac{a}{x} \coth (ax) \; .
\ee
Note that, in (\ref{imagecharge}), the combination that appears is again $\gamma L$. This
confirms our earlier result.

\section{Discussion}
\label{discussion}

We have seen that quotient spaces automatically break global Lorentz
invariance even when spacetime is precisely flat
everywhere. The key point is that the direction of identification
picks out a preferred frame and thereby makes it meaningful --- despite
the Minkowski metric --- to speak of the absolute velocity of objects. Individual inertial observers can then determine their state of motion by means of experiments that probe the global topology. In this paper we have pointed out that the exotic special-relativistic
effects that arise from the global breaking of Lorentz invariance have
a concrete realization in terms of brane worlds moving through flat compact
extra dimensions. Brane observers can potentially detect the motion of
the brane by experiments involving bulk fields, notably gravity. We
found three effects of brane motion: for large extra dimensions,
gravitons return to the brane with two sets of periodicities; for
small extra dimensions, there is a splitting of the Kaluza-Klein
spectrum; and in both cases there is an enhancement of the deviation
from the $1/r$ form of the Newtonian potential, which 
magnifies the extra dimensions by $\gamma$. There could well be other interesting effects.
Among the several questions this study raises is
the naturalness, or unnaturalness, of significant
brane velocity -- of significant $\gamma$. Perhaps brane gas cosmology establishes
a velocity distribution of branes that determines if a typical brane
would move at relativistic speeds.

We have made one simplifiying assumption: we neglected the
gravitational backreaction of the brane on the background
geometry. This assumption puts a constraint on the brane velocity and
brane tension,
which we can estimate heuristically. Consider a $3$-brane 
living in a
five-dimensional spacetime. From
the dimension-independent Poisson equation, $\nabla^2 V=4\pi G\rho$,
we can integrate over a spatial volume to get Gauss's Law, $- \int
f\cdot dA=4\pi G m_{\rm enc}$ relating the force per unit test mass to the
enclosed source mass. Consider a four-dimensional cube enclosing a piece of
the brane. The force per unit test mass goes as $f\propto G T_0$, where $T_0$ is the brane tension. The gravitational potential due to the brane therefore scales like
$V\propto GT_0 y$ as a function of distance $y$ in the extra
transverse direction. For the backreaction to be small, we require
$V \ll 1$. Now, the boosted energy density is $\gamma T_0$. Hence, with $G\propto \ell_P^{3}$, we find that the requirement that backreaction be small yields a condition on the product $\gamma T_o$ of the 3-brane
\be
G \gamma T_o L \ll 1 \Rightarrow \gamma \ll \frac{l_P}{L}\frac{1}{T_0 \, l_P^{4}}  \; .
\ee
Analogous considerations for $n$ extra dimensions ($n \geq 3$) indicate that backreaction can be neglected when
\be
G \gamma T_o \lf \frac{1}{L} \rt^{\! n-2} \ll 1 \Rightarrow \gamma \ll \lf \frac{L}{l_P} \rt ^{\! n-2} \frac{1}{T_0 \, l_P^{4}}  \; .
\ee
If the
brane tension is much lower than the Planck scale, the Lorentz factor
can be enormous.

Treated more formally, the presence of branes with tension poses
difficulties in a compact space. From the higher-dimensional point of
view, a brane is a delta-function source for the gravitational
field. In a compact space, however, there can be no net mass, just as
there can be no net Noether charge. (Gauss's Law gives inconsistent results: 
a positive source enclosed inside a cube versus zero source enclosed outside a cube.)
Thus the mass would have to be
canceled somehow. One could cancel it by adding a negative tension
brane, such as an orientifold. 
However, the presence of a second brane would interfere with the ability of gravitons to go around the extra dimension, if there were only one extra dimension. 
 A different possibility would be to consider only a
 single brane whose mass is canceled by a neutralizing background
 \cite{shtanov}. It would be interesting to find an explicit solution
 with brane motion through such a background, in which the background is still sufficiently close to flat so that Lorentz symmetry is still an approximate isometry.
These are technical difficulties, not physically prohibitive
obstacles. General relativity should allow for brane motion, even if the 
metric is resistant to analytic solutions. Still, genuine physical obstacles
could well interfere, as a
time-independent internal space might not be consistent with brane
motion for instance. It would be interesting to explore this, for
example by considering cosmology on a moving brane. It would also be
interesting to consider the effects of brane motion in the context of
specific phenomenologically viable brane world models.

\bigskip
\noindent
{\bf Acknowledgments}

\noindent
We would like to thank Gia Dvali, Pedro Ferreira, Lam Hui, Justin
Khoury, David Langlois, Edouardo Ponton, Massimo Porrati, Yuri
Shtanov, and Kishan Yerubandi for helpful discussions. 
JL gratefully acknowledges support of NSF
Theoretical Physics grant,
PHY - 0758022, and a KITP Scholarship.

\end{document}